%% file: kgons.tex
\documentclass[12pt]{article}

\usepackage{graphics}
\usepackage{latexsym}
\usepackage{color}
\usepackage{epsfig}

\usepackage{amsthm}
\usepackage{graphicx}
\usepackage{color}
\usepackage{amsmath}
\usepackage{amssymb}
\usepackage{amscd}
\usepackage{latexsym}
\usepackage{graphics}

\theoremstyle{plain}
\newtheorem{theorem}{Theorem}[section]
\newtheorem{lemma}[theorem]{Lemma}

\newtheorem{conjecture}[theorem]{Conjecture}
\newtheorem{observation}[theorem]{Observation}

\theoremstyle{definition}

\theoremstyle{remark}

\begin{document}

\title{Which $n$-Venn diagrams can be drawn with convex $k$-gons?}
\author{
Jeremy Carroll \and 
HP Laboratories, Bristol, UK \\
\and Frank Ruskey \and Mark Weston \\
University of Victoria, PO BOX 3055, Victoria, BC, Canada \\
}

\author{
Jeremy Carroll \footnotemark[2] \\
\and
Frank Ruskey \footnotemark[3] \\ 
\and
Mark Weston \footnotemark[3]\\
}

\maketitle                 
\renewcommand{\thefootnote}{\fnsymbol{footnote}}
\footnotetext[2]{ HP Laboratories, Bristol, UK}
\footnotetext[3]{ Department of Computer Science, PO BOX 3055, University of Victoria, Victoria, BC, Canada}
\renewcommand{\thefootnote}{\arabic{footnote}}

 \pagestyle{empty}

%

%
%

\begin{abstract}
We establish a new lower bound
for the number of sides required for the component curves
of simple Venn diagrams made from polygons.  Specifically, for any
$n$-Venn diagram of convex $k$-gons, we prove that $k \geq ( 2^n - 2 - n ) / ( n (n-2))$.
In the process we prove that Venn diagrams of seven curves, simple or not,
cannot be formed from triangles. 
We then give an example achieving the new lower bound of a (simple,
symmetric) Venn diagram of seven quadrilaterals.  Previously
Gr\"{u}nbaum had constructed a 7-Venn diagram of non-convex 5-gons 
[``Venn Diagrams II'', \emph{Geombinatorics} 2:25-31, 1992].
\end{abstract}

\section{Introduction and Background}

Venn diagrams and their close relatives, the Euler diagrams, form an
important class of combinatorial objects which are used in set theory,
logic, and many applied areas.  Convex polygons are fundamental geometric
objects that have been investigated since antiquity.
This paper addresses the question of which convex polygons can be used
to create Venn diagrams of certain numbers of curves.  This question has
been studied over several decades, for example~\cite{RRS, Au71, Gr75, Gr84, Gr92b}.
See the on-line survey~\cite{RW05} for more information on
geometric aspects of Venn diagrams.

Let $\mathcal{C} = \{ C_{1}, C_{2}, \ldots, C_{n} \}$ be a family of
$n$ simple closed curves in the plane.  The curves are required to be
finitely intersecting.  We say that $\mathcal{C}$ is a \emph{Venn diagram}
(or \emph{$n$-Venn diagram}) if all of the $2^n$ open regions
 $X_{1} \cap X_{2} \cap \cdots \cap X_{n}$ are non-empty and connected,
where each set $X_{i}$ is either the bounded interior or the unbounded
exterior of the curve $C_{i}$.  If the connectedness condition is dropped
the diagram is called an \emph{independent family}.  We can also think
of the diagram as a plane edge-coloured graph whose vertices correspond
to intersections of curves, and whose edges correspond to the segments
of curves between intersections. Edges are coloured according to the
curve to which they belong.  We can overload the term and also refer to
this graph as the \emph{Venn diagram}.
A Venn diagram or independent family is \emph{simple} if at most two curves intersect
at a common point, \emph{i.e.} every vertex has degree exactly four.  

Let the term \emph{$k$-gon} designate a \emph{convex} polygon with exactly $k$ sides.
Observe that two $k$-gons can (finitely) intersect with each other in at
most $2k$ points.  In this paper, we consider Venn diagrams and
independent families composed
of $k$-gons, for some $k$.  Note that the corners of the component $k$-gons
are not vertices in the graph interpretation of the diagram (unless
they intersect another curve at that point), and an edge, using the graph 
interpretation, may contain zero or more corners of the $k$-gon containing
that edge.  A \emph{side} of a $k$-gon is the line segment connecting two of its adjacent corners; sides
are not to be confused with edges.

It is clear that if
an $n$-Venn diagram can be drawn with $k$-gons, it can also be drawn with $j$-gons
for any $j > k$, by adding small sides and making small perturbations where necessary.

Gr\"{u}nbaum first considered the problem of what polygons can be used
to create Venn diagrams in~\cite{Gr75}, in which he gave a Venn diagram of
six quadrilaterals, and a diagram of five triangles.  He also provided
an independent family of five squares, and in~\cite{Gr92b} conjectured
that there is no symmetric Venn diagram with five squares.  

We restate two lemmas first observed by Gr\"{u}nbaum~\cite{Gr75},
some of the consequences of which inspired this work.   A \emph{FISC} is a family of \emph{F}initely 
\emph{I}ntersecting \emph{S}imple closed \emph{C}urves in the plane, with the property that 
the intersection of the interiors of all the curves is not empty.  Clearly, every Venn diagram is a FISC.

\begin{lemma}
In a FISC of $n$ convex $k$-gons there are at most ${ n \choose 2 } 2 k$ vertices.
\label{2k}
\end{lemma}

\noindent \textbf{Proof.} A pair of convex $k$-gons can intersect with each other at most $2k$ times; there
are ${n \choose 2}$ pairs.  \hfill $\Box$

\begin{lemma}
In a simple $n$-Venn diagram of $k$-gons, 
\begin{eqnarray} 
k & \geq & \left\lceil (2^{n-1} - 1) / { n \choose 2 } \right\rceil \mbox{ .} 
\end{eqnarray}
\label{gr}
\end{lemma}

\noindent \textbf{Proof.}  
Euler's formula for plane graphs, combined with the fact that in a simple diagram all vertices
have degree four, implies that the number of vertices in a simple Venn diagram is $2^{n} - 2$.
Combining this with Lemma~\ref{2k}, which gives an upper bound on
the number of vertices, the inequality follows.  \hfill $\Box$ 

\begin{figure}[h!tp]
\resizebox{\textwidth}{!}{
\includegraphics{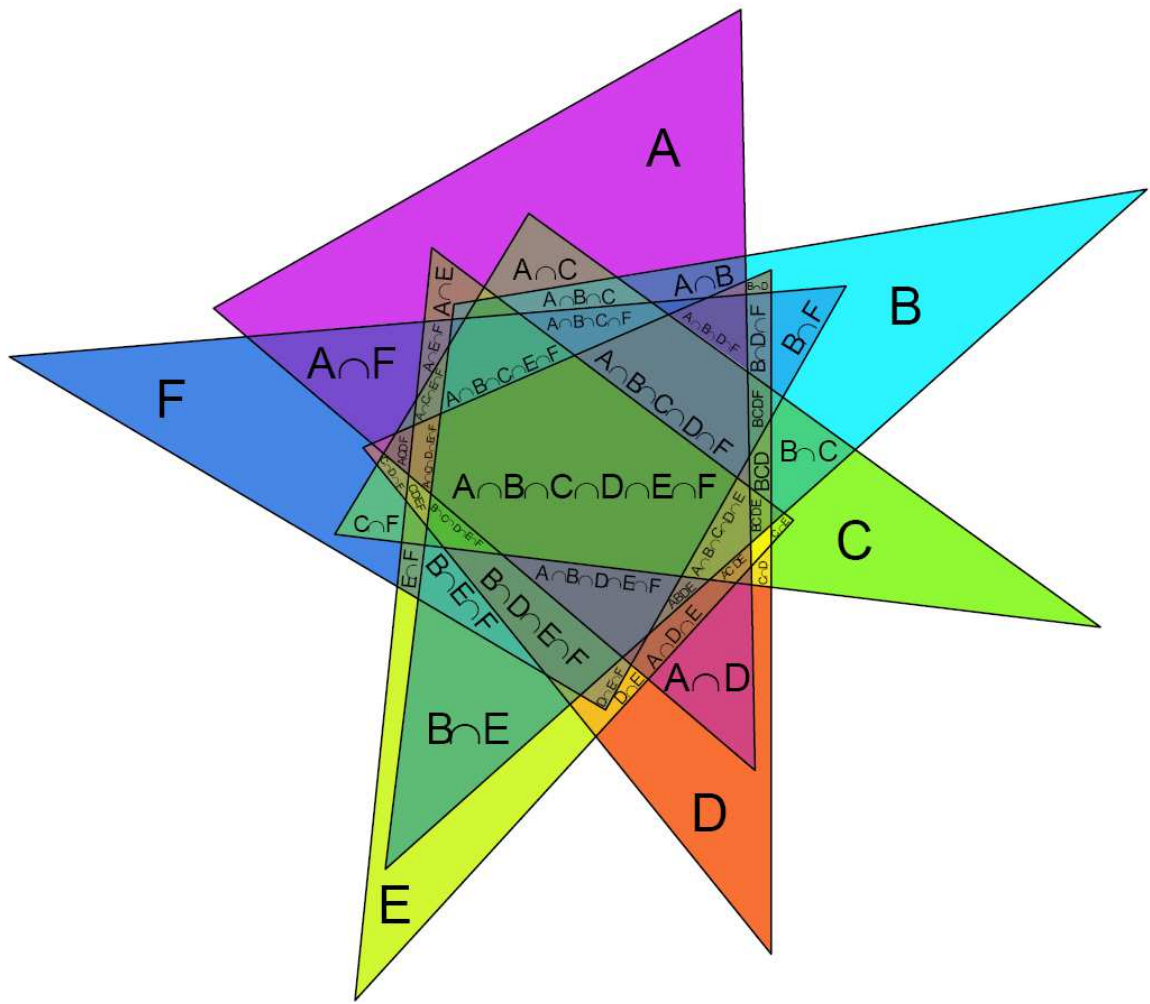} }
\caption{A Venn diagram of six triangles.}
\label{6triangles}
\end{figure}

\medskip
Lemma~\ref{gr} gives us a bound, for each $n$, of the minimum $k$ required
to form a simple $n$-Venn diagram of $k$-gons. 
Diagrams
are well-known that achieve the bounds for $n \leq 5$; see~\cite{RW05} for examples.
For $n = 6$, the Lemma implies
$k \geq 3$, and Carroll~\cite{Car00} achieved the lower bound by giving examples
of 6-Venn diagrams formed of triangles; his diagrams are all simple.  Figure~\ref{6triangles} shows one
of Carroll's
Venn diagrams of six triangles.  

For $n = 7$, Lemma~\ref{gr} implies that $k \geq 3$; however
until now the diagram with the smallest known $k$ was a 7-Venn diagram
of 5-gons by Gr\"{u}nbaum in~\cite{Gr92b}.  

The contributions of this paper are, first, to prove a tighter lower bound than Lemma~\ref{gr}
for the minimum $k$ required to draw a simple Venn diagram of $k$-gons; second,
to show that no 7-Venn diagram of triangles (simple or not)
can exist, and third, to achieve the new lower bound for $n=7$ by exhibiting a
Venn diagram of seven quadrilaterals.  

In~\cite{Gr75}, Theorem 3 contains bounds
on $k^*(n)$, which is defined as the minimal $k$ such that there exists a Venn diagram
of $n$ $k$-gons.  Carroll's results prove that $k^*(6) = 3$, and our results 
prove that $k^*(7) = 4$, and provide a lower bound on $k^*(n)$ for $n > 7$ when considering
simple diagrams.

\section{Venn Diagrams of $k$-Gons}

We now prove a tighter lower bound than that given by Lemma~\ref{gr} for simple Venn diagrams.  
Recall that a \emph{side} of a $k$-gon is the line connecting two of its adjacent corners, whereas
an \emph{edge} is the contiguous boundary of a $k$-gon between two adjacent vertices.
Vertices are points of intersection of two $k$-gons, as opposed to corners,
where two sides of a given $k$-gon meet.  An edge may contain zero or more corners of the $k$-gon containing it.

\begin{observation}
In a Venn diagram, each curve has exactly one edge on the outer face.
\label{outer}
\end{observation}

\noindent \textbf{Proof.} This is a special case of Lemma 4.6 from~\cite{CHP96}, which
states that no two edges in a face in a Venn diagram belong the same curves.  \hfill $\Box$

\bigskip

We now introduce some notation before proving the main theorems of this section.
In a Venn diagram of $k$-gons, consider any two $k$-gons $C_i$ and $C_j$, $1 \leq i < j \leq n$.  
The corners of $C_i$ may be
  labelled according to whether they are external ($E$) to $C_j$ or
  internal ($I$) to $C_j$ (we can assume that curves do not intersect
  at corners; if so, we can incrementally perturb one of the relevant curves to 
  eliminate this situation).  
In a clockwise walk around $C_i$ we obtain a circular sequence of
  $k$ occurrences of $E$ or $I$.
Let $\mathit{EI}_{ij}$ denote the number of occurrences of an $E$ label followed
  by an $I$ label; the notations $\mathit{II}_{ij}$ and $\mathit{IE}_{ij}$ are defined in an
  analogous manner.
We distinguish two cases when an $E$ follows an $E$: either
  $C_i$ is intersected twice on the side between the two $E$ corners,
  or it is not intersected.  
The notation $\mathit{EE}_{ij}$ is for the case where no intersection with
  $C_j$ occurs and $\mathit{EE'}_{ij}$ for the case where two intersections occur.
  By convexity, $C_i$ can only be intersected at most twice in a side by $C_j$.
Since these cases cover all types of corners,
\begin{equation}
\mathit{EI}_{ij} + \mathit{IE}_{ij} + \mathit{II}_{ij} + \mathit{EE}_{ij} + \mathit{EE'}_{ij} = k.
\label{1}
\end{equation}
Also note that $\mathit{EI}_{ij} = \mathit{IE}_{ij}$ since there must be an even number of crossings
between the curves.

\begin{theorem}
In a Venn diagram of $k$-gons, 
\begin{equation*}
V \leq 2k {n \choose 2} - n(k-1) \mbox{ . }
\end{equation*}
\label{Es}
\end{theorem}

\noindent \textbf{Proof. }

Given the notation above, consider the entire collection of curves.  Label corners on
  the outer face $\epsilon$ and the others $\iota$.
Define $E_i$ to be the number of corners of $C_i$ labelled $\epsilon$ and
  $I_i$ to be the number labelled $\iota$.
Clearly $I_i + E_i = k$.

In a Venn diagram each of the $n$ $k$-gons has one outer edge, by Observation~\ref{outer}, and
so all corners of $C_i$ labelled $\epsilon$ must appear
  contiguously; thus
\begin{equation}
\sum_{i \neq j} \mathit{EE}_{ij} \ge \sum_{i} (E_i -1)
\label{2}
\end{equation}
since the left-hand term will also count corners external to some curve but internal to others.

Since any corner labelled $\iota$ is internal to some curve,
\begin{equation}
\sum_{i \neq j} ( \mathit{II}_{ij} + IE_{ij} ) \ge \sum_{i} I_i
\label{3}
\end{equation}
since the left-hand term will double count any corner on $C_i$ internal to more than 1 curve.

Since each $\mathit{EI}$ and $\mathit{IE}$ accounts for one intersection and $\mathit{EE'}$ for two
  intersections,
\begin{eqnarray*}
2V & = & \sum_{i \neq j} ( \mathit{EI}_{ij} + \mathit{IE}_{ij} + 2 \mathit{EE'}_{ij} ) \\
   & = & \sum_{i \neq j} ( 2k - 2 \mathit{II}_{ij} - 2 \mathit{EE}_{ij} - \mathit{EI}_{ij} - \mathit{IE}_{ij} ) \hspace{10ex} \mbox{by (\ref{1}).} \\
   & = & 4k {n \choose 2} - \sum_{i \neq j} ( 2 \mathit{II}_{ij} + 2 \mathit{EE}_{ij} + 2 \mathit{EI}_{ij} ) \\
   & \le & 4k {n \choose 2} - 2 \sum_{i} (E_i -1) - 2 \sum_{i} I_i \hspace{10ex} \mbox{by (\ref{2}) and (\ref{3}).} \\
   & = & 4k {n \choose 2} - 2 \sum_{i} (E_i - 1 + I_i)  \\
   & = & 4k {n \choose 2} - 2 \sum_{i} (k - 1 )  \\
   & \le & 4k{n \choose 2} - 2n(k-1).
\end{eqnarray*}
Dividing by 2 gives $V \leq 2k {n \choose 2} - n(k-1)$, as desired. \hfill $\Box$

\bigskip

\begin{theorem}
In any simple $n$-Venn diagram of $k$-gons,
\begin{eqnarray*}
k & \geq & \left\lceil \frac{2^n - 2 - n }{n(n-2) } \right\rceil \mbox{ .}
\end{eqnarray*}
\label{simple}
\end{theorem}

\noindent \textbf{Proof. }
For simple Venn diagrams, we have that the number of vertices is $2^n - 2$.  Combined with Theorem~\ref{Es}, we have 
\begin{eqnarray*}
 2^n - 2 &  \leq &  2k {n \choose 2} - n ( k-1)  \\
&  = & n(n-1)k - nk + n \mbox{ . } \\
\mbox{Thus \hspace{10ex}} 2^n - 2 - n & \leq &  k( n(n-1) - n) \\
\mbox{ or \hspace{5ex} } \left\lceil \frac{2^n - 2 - n}{n(n-2)} \right\rceil & \leq &  k \mbox{, } 
\end{eqnarray*}
as desired.  \hfill $\Box$

\bigskip

Theorem~\ref{simple} gives a lower bound of the minimum $k$ required to 
construct a simple $n$-Venn diagram of $k$-gons.  Table~\ref{newbound} shows the bound
for small values of $n$.

For an upper bound on $k$, note that there are many general constructions
for Venn diagrams that produce diagrams of $k$-gons where the value $k$ is
a function of $n$ (for examples, see~\cite{PS} or~\cite{Gr75}).  In Gr\"{u}nbaum's convex
construction in~\cite{Gr75}, the $n$th curve is a
convex $2^{n-2}$-gon; this gives the upper bounds in Table~\ref{newbound} for $n > 7$.
Including this paper's contributions, diagrams are known for $n \leq 7$, thus solving these cases.

\begin{table}[h]
\begin{center}
\[ 
\begin{array}{c|cccccccccccc}
n      & 3  & 4  & 5 & 6  & 7  & 8  & 9   & 10  & 11  & 12   & 13   & 14 \\ \hline
k \geq & 1  & 2  & 2 & 3  & 4  & 6  & 8   & 13  & 21  & 35   & 58   & 98  \\
k \leq & 1  & 2  & 2 & 3  & 4  & 64 & 128 & 256 & 512 & 1024 & 2048 & 4096 \\ 
\end{array} 
\]
\end{center}
\caption{Minimum $k$ required to construct a simple $n$-Venn diagram of $k$-gons.}
\label{newbound}
\end{table}

\section{7-Venn Diagrams of Triangles}

In this section we prove that there is no 7-Venn diagram, simple or not,
composed of triangles.   The bound in Theorem~\ref{simple}, for $n=7$, gives
$k \geq 4$, which proves the simple case.

\begin{figure}
\input{split-nonseparable.pstex_t}
\caption{Splitting a vertex $v$ composed of intersecting straight-line segments.}
\label{split-fig}
\end{figure}
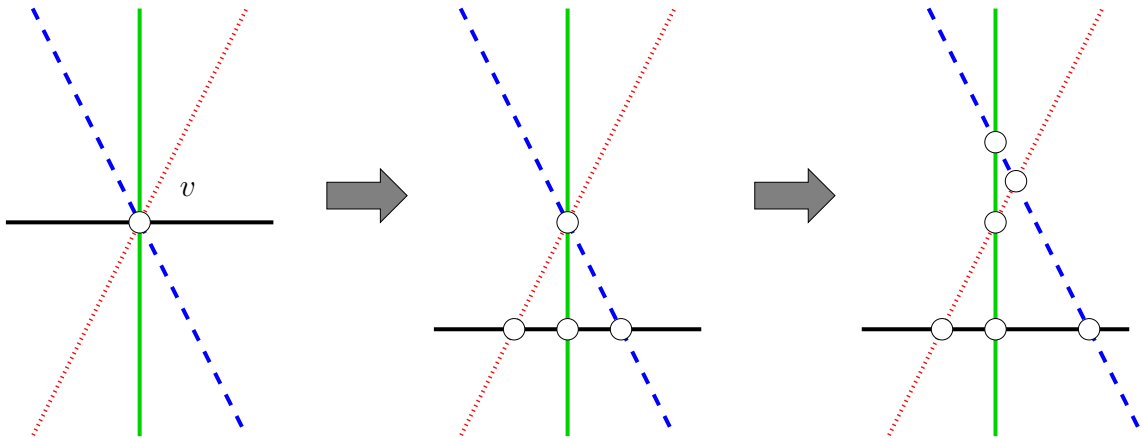

In a non-simple diagram, there must exist
at least one vertex where at least three curves intersect.  This vertex can
be reduced in degree by the operation of incrementally translating one
of the intersecting curves in a direction orthogonal to itself; this
operation will create a new face.  We call this operation \emph{splitting}
the vertex.  See Figure~\ref{split-fig} for an example; first the heavy
edge is translated, and then the dashed edge, reducing a vertex of degree
eight to six degree-four vertices, and three new faces are
created.

Note that splitting a vertex can never remove a face and
must add at least one face to the resulting diagram.  Thus, after
splitting any large degree vertex, the resulting diagram will no longer
be a Venn diagram as some face must be duplicated, but it will still be
an independent family.

Any large-degree vertex can be reduced to a set of vertices
of degree at most four by repeated splitting, as in the example above.  We
use this operation to prove the following:

\begin{lemma}
There is no non-simple Venn diagram of seven triangles.
\label{nonsimple}
\end{lemma}

\noindent \textbf{Proof.}
Assume such a diagram exists; call it $D_0$.  Since $D_0$ is non-simple,
some vertices have degree greater than four.  Let $D_1$ be the simple independent family formed by 
splitting all of the high-degree vertices in $D_0$.   
It is clear that this can be
performed while still retaining the fact that $D_1$ is composed of triangles, by incrementally
perturbing the corners of the component triangles.
Let $F_i$, $E_i$, and $V_i$ 
be the number of faces, edges, and vertices in $D_i$, for $i \in \{0,1\}$.

Since $D_0$ is a Venn diagram, $F_0 = 128$, and $F_1 > F_0$ since some new faces must have been
created by splitting vertices to form $D_1$.   Since $D_1$ has all degree-four vertices, 
summing the vertex degrees gives us $E_1 = 2 V_1$.   Using Euler's formula, $V_1 + F_1 
- E_1 = 2$, and substituting for $E_1$ gives $V_1 = F_1 - 2 > F_0 - 2 = 126$, and so
$V_1 > 126$.  

However, $D_1$ is composed of triangles, two of which can only
intersect at most six times, and thus $V_1 \leq 6 { 7 \choose 2 } = 126$, which
provides a contradiction.
\hfill $\Box$

\medskip

\begin{theorem}
There is no Venn diagram of seven triangles.
\label{7triangles}
\end{theorem}

\noindent
\textbf{Proof.} By Theorem~\ref{simple} and Lemma~\ref{nonsimple}. \hfill $\Box$

\section{7-Venn Diagrams of Quadrilaterals}

The proof of the previous section raises the question of how close we can get: is there
a Venn diagram constructed of seven 4-gons?  In this section we answer this question in the affirmative with
a simple diagram; this shows that the bound in Theorem~\ref{simple} is tight for $n \leq 7$.

\begin{figure}[h!tp]
\resizebox{\textwidth}{!}{
\includegraphics{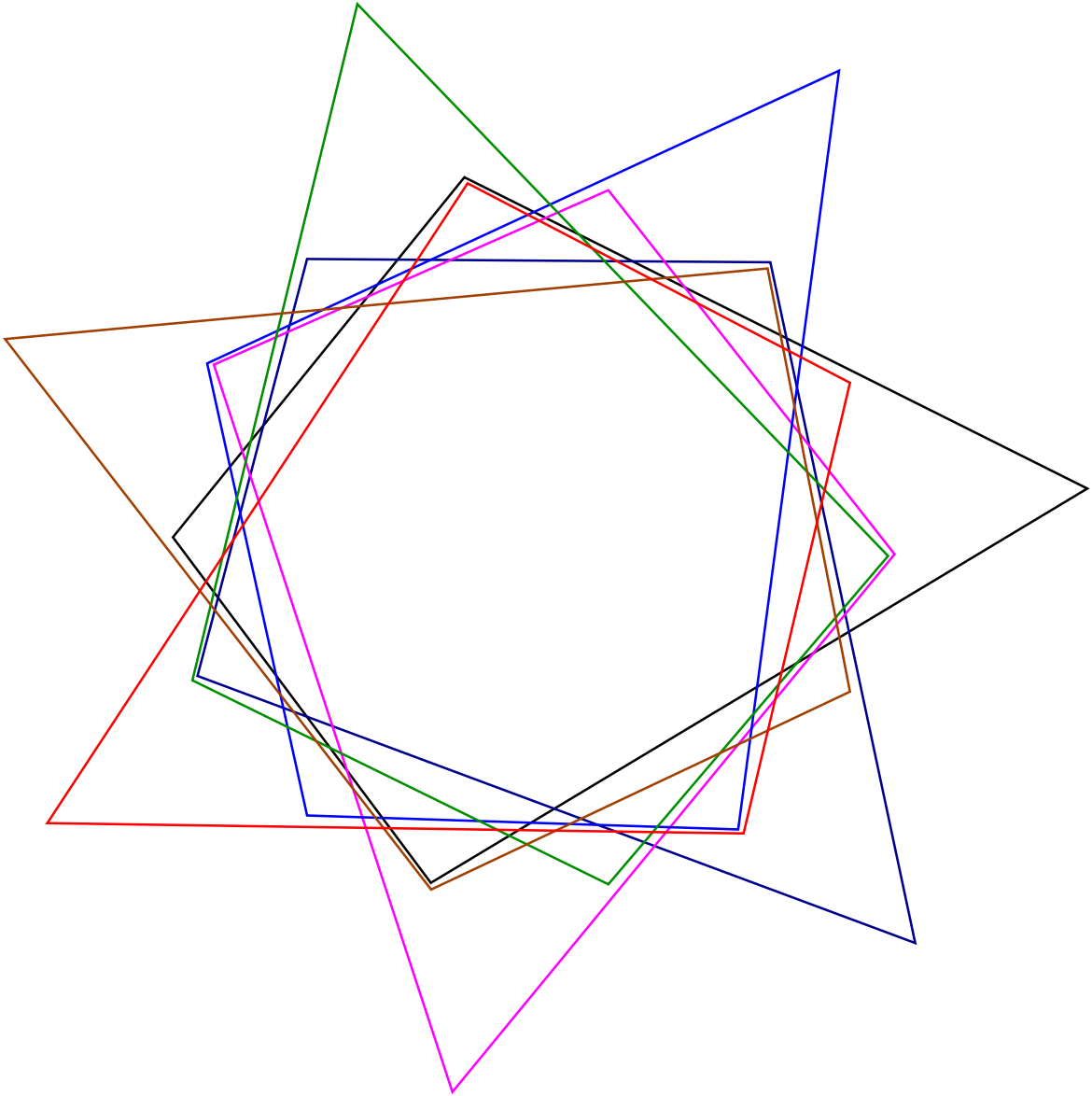} }
\caption{A (symmetric, simple) Venn diagram of seven quadrilaterals.}
\label{7quads}
\end{figure}

Figure~\ref{7quads} shows a simple 7-Venn diagram of quadrilaterals.
The diagram is also \emph{symmetric}: it possesses a rotational symmetry
about a centre point, and the seven quadrilaterals
are congruent as each maps onto the next by a rotation of $2 \pi / 7$.
The figure is in fact isomorphic as a plane graph (\emph{i.e} can
be transformed by continuous deformation in the plane) to the 7-Venn
diagram ``Victoria'' discovered by Frank Ruskey~\cite[``Symmetric Venn
Diagrams'']{RW05}.

Table~\ref{coords} gives the coordinates for the four corners of one
of the component quadrilaterals: the other six quadrilaterals may be
constructed by rotating the given coordinates around the origin by an
angle of $2 \pi i / n$, for $1 \leq i \leq 6$.

\begin{table}[h]
\begin{center}
\[ 
\begin{array}{r@{,\ }l}
(x& y) \\ \hline
(-0.446 & \ \ 0.000 ) \\
(-0.123 & -0.433 ) \\
(\ \ 0.699 & \ \ 0.061 ) \\
(-0.081 & \ \ 0.451 )
\end{array} 
\]
\end{center}
\caption{Coordinates of corners of a quadrilateral in Figure~\ref{7quads}.}
\label{coords}
\end{table}

This diagram was discovered using a simple software tool to manipulate polygons in the
plane and compute intersections between them.


\section{Open Problems}

It is not known whether the bound in Theorem~\ref{simple} is tight for $n > 7$.
Note that the non-simple
result in Lemma~\ref{nonsimple} works because of the fortuitous fact that $2^n - 2 = {n \choose 2} 2k$ for 
$n=7$ and $k = 3$, which is not true for $n \geq 8$.  Thus this technique will not work for establishing a non-simple
lower bound for $n \geq 8$.  A nice open problem is thus to find a tight lower bound on $k$ for the existence of 
simple and non-simple $n$-Venn diagrams of $k$-gons in general.  It appears difficult to generalize Theorem~\ref{simple}
to the non-simple case; nevertheless we offer: 

\begin{conjecture}
The bound in Theorem~\ref{simple} also holds for non-simple diagrams.
\end{conjecture}

Not all 7-Venn diagrams can be drawn with quadrilaterals; for example, in the diagram
M4 from~\cite[``Symmetric Venn Diagrams'']{RW05}, each curve has another curve intersect with it 
10 times, implying that at least 5-gons are required to draw the figure with $k$-gons.
Thus, what is the maximum over all $n$-Venn diagrams
of the minimum $k$ required to draw each diagram as a collection of $k$-gons?

\section{Acknowledgements}

Many thanks to Branko Gr\"{u}nbaum for helpful discussions about the
topics of this paper. 

\bibliographystyle{plain}

\bibliography{kgons}

\end{document}

%% file: split-nonseparable.pstex_t
\begin{picture}(0,0)%
\includegraphics{split-nonseparable.pstex}%
\end{picture}%
\setlength{\unitlength}{2210sp}%
\begingroup\makeatletter\ifx\SetFigFont\undefined%
\gdef\SetFigFont#1#2#3#4#5{%
  \reset@font\fontsize{#1}{#2pt}%
  \fontfamily{#3}\fontseries{#4}\fontshape{#5}%
  \selectfont}%
\fi\endgroup%
\begin{picture}(12837,4888)(3857,-9480)
\put(5851,-6736){\makebox(0,0)[lb]{\smash{{\SetFigFont{12}{14.4}{\rmdefault}{\mddefault}{\updefault}{\color[rgb]{0,0,0}$v$}%
}}}}
\end{picture}%